\documentclass{cimento}
\usepackage{caption}
\usepackage{amsmath}
\usepackage{amssymb}
\usepackage{amsfonts}
\usepackage{graphicx}
\usepackage{pstricks}
\title{Error analysis of the parameters of the Skyrme N2LO pseudo-potential}
\author{P.~Becker\from{york}\ETC,
A.~Pastore\from{york},
D.~Davesne\from{lyon},
        \atque
J.~Navarro\from{ific}}
\instlist{\inst{york} Department of Physics, University of York, Heslington, York, Y010 5DD, United Kingdom
  \inst{lyon} Universit\'e de Lyon, Universit\'e Lyon 1, 43 Bd. du 11 Novembre 1918, F-69622 Villeurbanne cedex, France\\
             CNRS-IN2P3, UMR 5822, Institut de Physique Nucl{\'e}aire de Lyon 
             \inst{ific} IFIC (CSIC-Universidad de Valencia), Apartado Postal 22085, E-46.071-Valencia, Spain }

\begin{document}

\maketitle

\begin{abstract}
We perform a detailed analysis of the parameters of the Skyrme SN2LO interaction. By mean of a covariance matrix analysis, we have been able to provide error bars and build the complete covariance matrix. The latter will be used in future applications to properly study the propagation of errors on nuclear observables calculated with this pseudo-potential.
\end{abstract}

\section{Introduction}

Nuclear Energy Density Functional (NEDF) theory is the only microscopic model able to reproduce both ground state and excited state properties of atomic nuclei from drip-line to drip-line and from light to super-heavy nuclei~\cite{erl12,ben03}. Among the different functionals, the ones generated by the Skyrme interaction~\cite{sky59} are among the most popular and most successful in describing properties of atomic nuclei~\cite{gor09,sab07}.

Once the functional form is fixed, one needs to determine the value of the effective parameters using some selected nuclear observables. The quality of the predictions made with such a functional strongly depends on the accuracy of such an adjustment. In a recent series of articles~\cite{une0,une1,une2}, the UNEDF collaboration~\cite{bog13} has explored the flexibility of the standard Skyrme functional to understand whether the observed discrepancy with experimental data may or may not be reabsorbed into the coupling constants of the functional by using more refined fitting protocols.
The outcome of their last article~\cite{une2} is that the Skyrme functional has reached its limits and it is mandatory to explore either other functional forms or different many-body methods. 

Inspired by the previous work of Refs.~\cite{car08,rai11}, we have extended the original Skyrme pseudo-potential~\cite{sky59}, including higher order momentum terms~\cite{dav13}.  In Refs.~\cite{dav15,dav16}, we have shown that such higher order terms mimic the physics of an effective range by including explicit contributions of higher order partial waves. In his original article~\cite{sky59}, Skyrme found that \emph{any} finite range interaction can be expanded in momentum space and that with the incorporation of only S and P waves, the description of nuclear data would be satisfactory. However, he also mentioned in the same paper that the inclusion of the D wave term might have a non negligible role in low-energy nuclear structure.
Similar conclusions have been also obtained in a more recent article~\cite{car10} within the context of Density Matrix Expansion.

In Ref.~\cite{bec17}, we have been able to provide for the very first time a stable parametrisation of the \emph{extended} Skyrme pseudo-potential, hereafter called SN2LO1, by performing a fit on atomic nuclei with an improved version of Saclay-Lyon fitting protocol~\cite{pas13,cha98}. In the present article, we aim at continuing the investigation, by performing a series of statistical tests to quantify the goodness of the fit, but also to quantify error bars on the parameters we have obtained.This aspect is very important for a future rigorous treatment of errors. See Ref.~\cite{dob14} for more details.

The article is organised as follows: in Sec.~\ref{sec:n2lo}, we briefly discuss the main feature of the extended Skyrme pseudo-potential. In Sec.~\ref{sec:fit} we recall the fitting protocol used to fix the parameters of the pseudo-potential, while in Sec.~\ref{sec:cov} we perform a rigorous statistical analysis of the results. We present our conclusions in Sec.~\ref{sec:concl}.

\section{Skyrme N2LO}\label{sec:n2lo}

The N2LO Skyrme pseudo-potential, as described in Refs.~\cite{car08,rai11,dav13}, is a generalisation of the standard Skyrme interaction, corresponding to the expansion of a generic finite-range interaction in powers of the relative momenta $\mathbf{k}, \mathbf{k}'$ up to the fourth order. It is written as the sum of three terms: central, spin-orbit and density dependent~\cite{dav16}
\begin{equation}\label{eq:N2LO}
V_{\text{N2LO}} =V_{\rm N2LO}^{C}+V_{\rm N1LO}^{LS}+V_{\rm N1LO}^{DD}\;.
\end{equation}

\noindent The central term reads
\begin{eqnarray} \label{eq:N2LO:c}
V_{\rm N2LO}^{C} &=& t_0 (1+x_0 P_{\sigma}) + \frac{1}{2} t_1 (1+x_1 P_{\sigma}) ({\mathbf{k}}^2 + {\mathbf k'}^2)   + t_2 (1+x_2 P_{\sigma}) ({\mathbf k} \cdot {\mathbf k'})  \nonumber\\
            & & + \frac{1}{4} t_1^{(4)} (1+x_1^{(4)} P_{\sigma}) \left[({\mathbf k}^2 + {\mathbf k'}^2)^2 + 4 ({\mathbf k'} \cdot {\mathbf k})^2\right] \nonumber\\
            &&+ t_2^{(4)} (1+x_2^{(4)} P_{\sigma}) ({\mathbf k'} \cdot {\mathbf k}) ({\mathbf k}^2 + {\mathbf k'}^2) .
\end{eqnarray}
In the above expression, a Dirac function $\delta({\mathbf r}_1-{\mathbf r}_2)$ is to be understood~\cite{ben03}. For the spin-orbit term $V_{\rm N1LO}^{LS}$, we  have used
\begin{eqnarray}
V_{\rm N1LO}^{LS}&=&i W_0 (\hat{\sigma}_1+\hat{\sigma}_2)\left[{\mathbf{k}}'\times  {\mathbf{k}} \right]\;,
\end{eqnarray}

 \noindent as in the standard Skyrme interaction~\cite{cha98}. In Ref.~\cite{dav16}, we have shown that a finite-range spin-orbit term can be also expanded in terms of relative momenta, but only this one is actually gauge invariant and thus fulfils a continuity equation~\cite{rai11}.
For the density dependent term we have adopted the standard term as used in Ref.~\cite{cha98}.
%
%
 A possible alternative would be the inclusion of an explicit three-body term. This possibility has been discussed in details in Ref.~\cite{sad13}. Let us finally mention that from the above expressions, it is possible to derive the Skyrme N2LO functional by averaging on Hartree-Fock states (HF). We refer to Ref.~\cite{bec17} for more details.

 \section{Fitting protocol}\label{sec:fit}

The  14 parameters $\mathbf{p}=\{t_0,t_1,t_1,\dots\}$ of the N2LO pseudo-potential are determined by minimising a penalty function containing both properties of the infinite nuclear medium and finite nuclei.
The penalty function $\chi^2$ is defined as~\cite{dob14}

\begin{eqnarray}\label{chi2}
\chi^2(\mathbf{p})=\sum_{i=1}^M \left( \frac{ \mathcal{O}_i-f_i(\mathbf{p})}{\Delta \mathcal{O}_i}\right)^2
\end{eqnarray} 

\noindent where the sum runs over all the M (pseudo)-observables $\mathcal{O}_i$ we include in the fit. $f_i$ is the value obtained with our model for a given array of parameters $\mathbf{p}$ of size $n$. Finally $\Delta \mathcal{O}_i$ is the weight we give to each point in the fit. For more details see Ref.~\cite{bec17}.
The fitting protocol is quite similar to the usual Saclay-Lyon one~\cite{cha98} combined with the nuclear response functions in infinite nuclear matter~\cite{pas15} to avoid all possible finite-size instabilities~\cite{hel13}. To this purpose we had to add on top of Eq.~(\ref{chi2}) an additional asymmetric constraint~\cite{bec17}.
The resulting parametrisation, named SN2LO1, fairly reproduces binding energies, radii~\cite{bec17} and pairing gaps~\cite{bec18} with the same level of accuracy of the SLy5* functional~\cite{pas13}, fitted with the same protocol. However, even if it is the first time a stable parametrisation incorporating higher order gradients have been obtained, the higher order interaction parameters are actually not well constrained and a major challenge is to find the right observables to constraint them. 

\section{Covariance analysis}\label{sec:cov}

The minimisation of Eq.\ref{chi2} leads to the optimal set of parameters. Let us call it $\mathbf{p}_0$. Following Ref.~\cite{roc15}, we can thus make a Taylor expansion of $\chi^2$ around $\mathbf{p}_0$ as

\begin{eqnarray}\label{taylor}
\chi^2(\mathbf{p})=\chi^2(\mathbf{p}_0)+\frac{1}{2}\sum_{ij}^n(p_i-p_{0i})\left. \frac{\partial^2\chi^2(\mathbf{p}) }{\partial p_i \partial p_{j}}\right|_{\mathbf{p}=\mathbf{p}_{0}}(p_j-p_{0j})+\dots
\end{eqnarray}

\noindent The covariance matrix $\mathcal{E}$ is defined as the inverse of the Hessian matrix $\mathcal{M}$
\begin{eqnarray}
\mathcal{M}_{ij}&=&\left.\frac{\partial^2\chi^2(\mathbf{p}) }{\partial p_i \partial p_{j}}\right|_{\mathbf{p}=\mathbf{p}_{0}}\;.
\end{eqnarray}

\noindent We immediately see from this equation that the derivatives with respect to the parameters need to be computed numerically thus requiring an optimal choice for the variation step of the parameters. In the present work, we have adopted the criterion given in Ref.~\cite{roc15}.
From  $\mathcal{E}$, we then extracted the error bar on the fitted parameter $p_i$ as $e(p_i)=\sqrt{\mathcal{E}_{ii}}$
together with some possible correlations between parameters
\begin{eqnarray}
\mathcal{C}_{ij}=\frac{\mathcal{E}_{ij}}{\sqrt{\mathcal{E}_{ii}\mathcal{E}_{jj}}}\;,
\end{eqnarray}
$\mathcal{C}$ being the correlation matrix. More precisely, when $\mathcal{C}_{ij}=\pm1$ then the two parameters $p_i,p_j$ are correlated (+1) or anti-correlated (-1). When  $\mathcal{C}_{ij}\approx0$ the two parameters are independent. See Ref.~\cite{dob14} for a more detailed discussion.


In performing our analysis, we have observed that some of the parameters of our SN2LO1 functional were poorly constrained showing a very flat $\chi^2$ in some directions in parameter space. The resulting functional is \emph{sloppy}~\cite{nik16}. As a consequence a full covariance analysis would be meaningless, 
the optimal solution being to find observable(s) for which the new terms $t_1^{(4)},t_2^{(4)}$ are most sensitive. Current sensitivity studies are ongoing, but at present we have not been able to find adequate observable(s).
As a consequence and an alternative, we have decided to keep the same fitting protocol of Ref.~\cite{bec17}, but now modifying slightly the weights of the different components of the penalty function $\chi^2$ as done in Ref.~\cite{was12}.
Changing the weights $\Delta\mathcal{O}$ effectively modifies the hyper-surface of the penalty function and gives us the possibility of exploring other local minima.
We have thus been able to find a physically acceptable local minimum where the parameters of the Skyrme functional are all well-constrained~\cite{dob14}.
In Tab.~\ref{tab:inter}, we report the parameters with their corresponding error bar of the new pseudo-potential named SN2LO1$_\text{cov}$.

\begin{center}
\begin{tabular}{cc|cc}
\hline
\hline
$n$ & $i$ & $t_i^{(n)}$ [MeVfm$^{3+n}$] & $x_i^{(n)}$\\
\hline
 0 & 0 &-2477.48 $\pm$ 5.87&0.766 $\pm$0.014\\
 2 & 1 &  507.80 $\pm$ 6.64 & 0.001$\pm$0.003\\
 2 & 2& -421.48 $\pm$  23.07& -0.918 $\pm$ 0.012\\
4 & 1 & -29.68 $\pm$4.67&  0.352  $\pm$0.006\\
4 & 2 &  4.69$\pm$ 0.63&2.14 $\pm$0.057\\
\hline
\multicolumn{4}{c}{$t_3 [\mbox{MeV fm}^{3(1+\alpha)}]$ =13660.67 $\pm$54.94  $x_3= 1.173\pm0.018$}\\
\hline
\multicolumn{4}{c}{$W_0$  = 122.3 $\pm$ 6.7  [MeV fm$^5$]}\\
\hline
\hline
\end{tabular}
\captionof{table}{Parameters of the SN2LO1$_{\text{cov}}$ pseudo-potential and their associated error bar. We keep the exponent $\alpha=1/6$ fixed during the fit.}
\label{tab:inter}
\end{center}

\begin{figure}[!h]
\begin{center}
\includegraphics[width=0.58\textwidth]{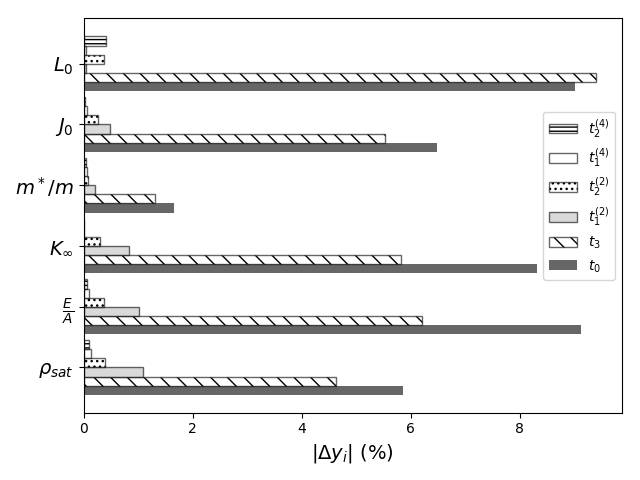}
\end{center}
\caption{
Propagation of a 1\% variation of a few selected parameters of SN2LO1$_{\text{cov}}$ on some properties of nuclear matter at saturation density $\rho_{sat}$. See text for details. }
\label{fig:variation}
\end{figure}

By inspecting the above table, we observe that all parameters are tightly constrained and that the new higher order terms, although very small, are not compatible with zero. This means that these parameters are not redundant and they may used to improve the fit. We may however notice that the $x_1^{(2)}$ parameter is compatible with 0, illustrating that the higher order parameters lead to a non-perturbative reorganisation of some of the previous Skyrme NLO parameters.
In Fig.~\ref{fig:variation}, we study the propagation of a 1\% variation on the $t_i^{(n)}$ parameters on some selected properties of nuclear matter at saturation density $\rho_{\text{sat}}$ : symmetry energy $J_0$ and its slope $L_0$, effective mass $m^*/m$, binding energy per particle $E/A$ and nuclear incompressibility $K_{\infty}$. As expected, the major contributions come from the $t_0,t_3$ terms, which are very tightly constrained in our fit, the terms $t_{1}^{(4)},t_{2}^{(4)}$ having a non negligible impact only on the slope of the symmetry energy.

%
%


\begin{figure}[!h]
\begin{center}
\includegraphics[width=0.6\textwidth]{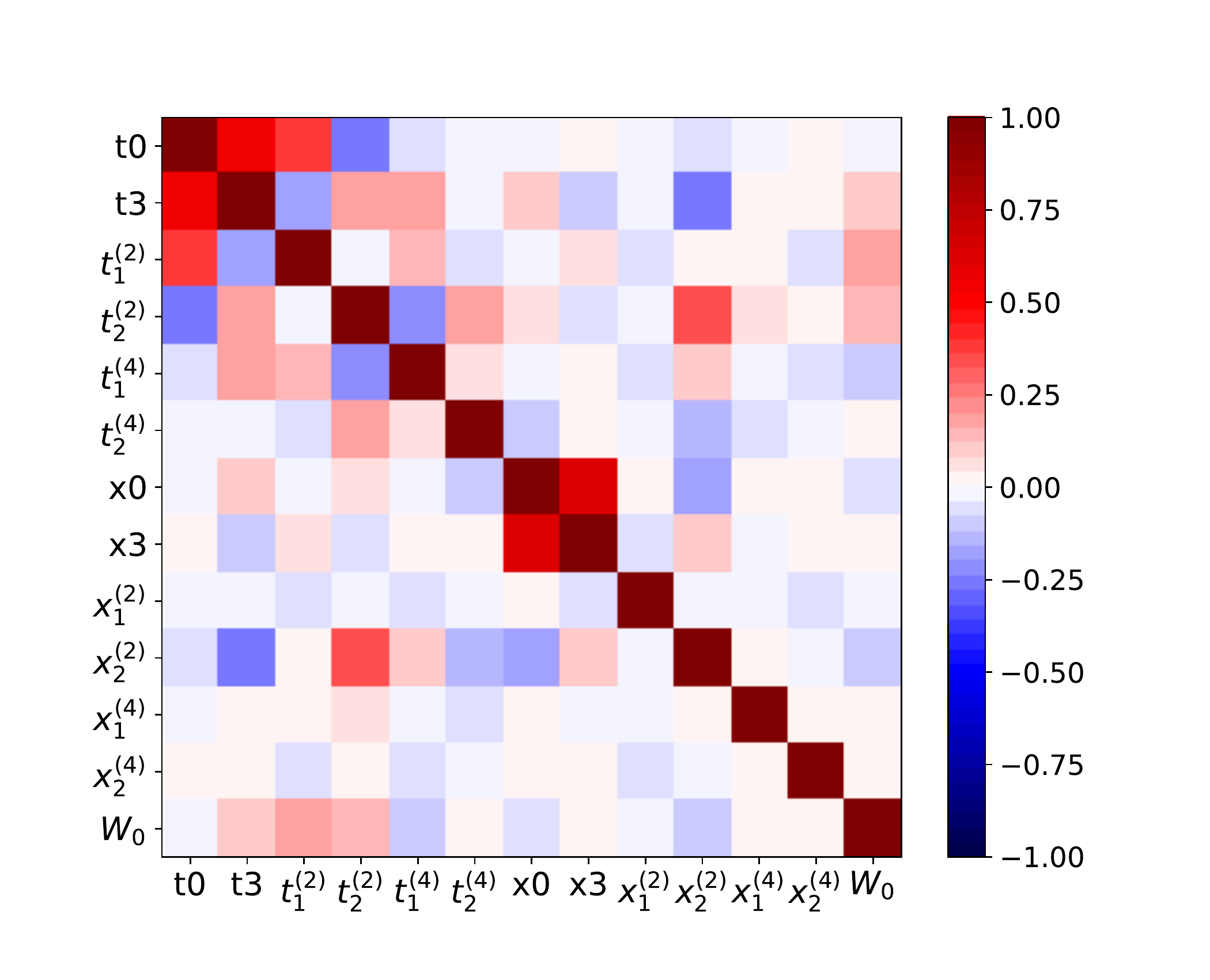}
\end{center}
\caption{(Colors online) Correlation matrix for SN2LO1$_{\text{cov}}$ parameters. See text for details. }
\label{corr:mat}
\end{figure}

In Fig.~\ref{corr:mat}, we also show the correlation matrix $\mathcal{C}$ for the SN2LO1$_{\text{cov}}$ parametrisation.  
We observe how  the $t_i$ parameters are on average strongly correlated among each other. For example we see that the $t_0,t_3$ parameters are strongly correalted with the $t_1^{(2)}$; while the fourth order terms are correlated to their corresponding second order one in pairs: $t_1^{(4)}/t_1^{(2)}$ and $t_2^{(4)}/t_2^{(2)}$. In Ref.~\cite{dav18}, we have shown the strong correlation between the effective mass and other saturation properties of the nuclear medium and thus effectively a strong correlation between the $t_i$ parameters.
An interesting feature of Fig.~\ref{corr:mat} is that the $x_i$ parameters are quite independent from each other, apart for $x_0$ and $x_3$.
This means we may have some extra freedom in adjusting them in a future fit.

\section{Conclusions}\label{sec:concl}

We have performed a full covariance analysis on the recently fitted Skyrme N2LO pseudo-potential~\cite{bec17}. By introducing some modifications to the weights of the penalty function, we have been able to find a well defined local minimum to perform a full covariance analysis.
We have derived error bars for the parameters of the pseudo-potential and we have noticed that the new parameters are non-compatible with zero, thus showing that these new terms do actively contribute to the fit.

At present, it is still not possible to judge if these new parameters will lead to a real improvement in describing nuclear observables, but research in this direction is ongoing.

\acknowledgments
The work of J.N. has been supported by grant FIS2017-84038-C2-1-P, Mineco (Spain).
The work of A.P. is supported  by the UK Science and Technology Facilities Council under Grants No. ST/L005727 and ST/M006433.


\begin{thebibliography}{0}
\bibitem{erl12}\BY{Erler J. \etal} \IN{Nature}{486}{2012}{509}
\bibitem{ben03}\BY{Bender M., Heenen~P.-H \atque  Reinhard~P.G.} \IN{Rev. Mod. Phys.}{75}{2003}{121}
\bibitem{sky59} \BY{Skyrme T.H.R.} \IN{Nucl. Phys}{9}{1959}{615}
\bibitem{gor09} \BY{Goriely S., Chamel N. \atque Pearson J.M.} \IN{Phys. Rev. Lett}{102}{2009}{152503}
\bibitem{sab07} \BY{Sabbey B., Bender M., Bertsch G.F. \atque Heneen P.H.} \IN{Phys. Rev. C}{75}{2007}{044305}
\bibitem{une0}\BY{Kortelainen M. \etal} \IN{Phys. Rev. C}{82}{2010}{024313}
\bibitem{une1}\BY{Kortelainen M. \etal} \IN{Phys. Rev. C}{85}{2012}{024304}
\bibitem{une2}\BY{Kortelainen M. \etal} \IN{Phys. Rev. C}{89}{2014}{054314}
\bibitem{bog13}\BY{Bogner S. \etal} \IN{Comp. Phys. Comm.}{184}{2013}{2235-2250}
\bibitem{car08} \BY{Carlsson B.G., Dobaczewski J. \atque Kortelainen M.} \IN{Phys. Rev. C}{78}{2008}{044326}
\bibitem{rai11}\BY{Raimondi F.,  Carlsson B.G. \atque Dobaczewski J.} \IN{Phys. Rev. C}{83}{2011}{054311}
\bibitem{dav13}\BY{Davesne D., Pastore A. \atque  Navarro J.} \IN{J. Phys. G}{40}{2013}{095104}
\bibitem{dav15}\BY{Davesne D., Holt J.W., Pastore A. \atque Navarro J.} \IN{Phys. Rev. C}{91}{2015}{064303}
\bibitem{dav16}\BY{Davesne D., Becker P., Pastore A. \atque Navarro J.} \IN{Ann. Phys.}{375}{2016}{288-312}
\bibitem{car10}\BY{Carlsson B. G \atque  Dobaczewski~J.} \IN{Phys. Rev. Lett.}{105}{2010}{122501}
\bibitem{bec17} \BY{Becker P., Davesne D., Meyer J., Navarro J. \atque Pastore A.} \IN{Phys. Rev. C}{96}{2017}{044330}
\bibitem{pas13}\BY{Pastore A., Davesne D., Bennaceur K., Meyer J. \atque Hellemans V.} \IN{Phys. Scripta}{T154}{2013}{014014}
\bibitem{cha98}\BY{Chabanat E., Bonche P., Haensel P., Meyer J.,\atque Schaeffer R.} \IN{Nucl. Phys. A}{635}{1998}{231-256}
\bibitem{dob14}\BY{Dobaczewski J., Nazarewicz W. \atque Reinhard P.G.} \IN{J. Phys. G}{41}{2014}{074001}
\bibitem{sad13}\BY{Sadoudi J., Duguet T., Meyer J. \atque Bender M.} \IN{Phys. Rev. C}{88}{2013}{064326}

\bibitem{pas15} \BY{Pastore A., Davesne D. \atque Navarro J.} \IN{Phys. Reports}{563}{2015}{1-67}
\bibitem{hel13} \BY{Hellemans V \etal} \IN{Phys. Rev. C}{88}{2013}{064323}
\bibitem{bec18} \BY{Becker P., Davesne D., Meyer J., Navarro J. \atque Pastore A.} \IN{Acta Phys. Pol. B}{49}{2018}{331}
\bibitem{roc15} \BY{Roca-Maza X., Paar N. \atque Col\'o G} \IN{J. Phys. G}{42}{2015}{034033}
\bibitem{tia}\BY{Haverinen T. \atque Kortelainen M.} \IN{J. Phys. G}{44}{2017}{044008}
\bibitem{nik16}\BY{Niksic T. \atque Vretenar D.} \IN{Phys. Rev. C}{94}{2016}{024333}

\bibitem{was12} \BY{Washiyama K., Bennaceur K., Avez B., Bender M., Heenen P.H. \atque Hellemans V.} \IN{Phys. Rev. C}{86}{2012}{054309}
%
\bibitem{dav18}\BY{Davesne D., Navarro J., Meyer J., Bennaceur K. \atque Pastore A.} \IN{Phys. Rev. C}{97}{2018}{044304}


\end{thebibliography}
\end{document}